# ELASTICITY: Topological Characterization of Robustness in Complex Networks

A. Sydney, C. Scoglio, P. Schumm
EECE
Kansas State University
{asydney, caterina, pbschumm}@ksu.edu

R. E. Kooij
Delft University of Technology
TNO ICT the Netherlands
robert.kooij@tno.nl

## ABSTRACT
Just as a herd of animals relies on its robust social structure to survive in the wild, similarly robustness is a crucial characteristic for the survival of a complex network under attack. The capacity to measure robustness in complex networks defines a network's survivability in the advent of classical component failures and at the onset of cryptic malicious attacks. To date, robustness metrics are deficient and unfortunately the following dilemmas exist: accurate models necessitate complex analysis while conversely, simple models lack applicability to our definition of robustness. In this paper, we define robustness and present a novel metric, elasticity- a bridge between accuracy and complexity-a link in the chain of network robustness. Additionally, we "test-drive" the performance of elasticity on Internet topologies and online social networks, and articulate results. *

## 1. INTRODUCTION
The gravity of network robustness requires earnest attention. In 2001, Code Red, a computer virus that incapacitated numerous networks, resulted in a global loss of 2.6 billion US dollars. In 2004, the Sassar virus caused Delta airlines to cancel 40 transatlantic flights in addition to halting trains in Australia. The US General Accounting Office estimated 250,000 annual attacks on Department Of Defense networks. Objectives of such attacks range from theft, modification, and destruction of data to dismantling of entire networks. Our daily routines would cease to exist should the technological information infrastructure disintegrate. Thus, it becomes crucial to maintain the highest levels of robustness in complex networks.

In an effort to measure robustness, it is necessary to first provide a definition for the robustness. Amongst other definitions, a network can be robust if disconnecting components is difficult. However, in this paper, we define a robust network as one where the total throughput degrades gracefully under node and link removal. The former definition is based on topological characteristics, while the latter also considers traffic flow.

The classical approach for determining robustness of networks entails the use of basic concepts from graph theory. For instance, the connectivity of a graph is an important, and probably the earliest, measure of robustness of a network [1]. Node (link) connectivity, defined as the size of the smallest node (link) cut, determines in a certain sense the robustness of a graph to the deletion of nodes (links). However, the node or link connectivity only partly reflects the ability of graphs to retain certain degrees of connectedness after deletion. Other improved measures were introduced and studied, including super connectivity [2], conditional connectivity [3], restricted connectivity [4], fault diameter [5], toughness [6], scattering number [7], tenacity [8], expansion parameter [9] and isoperimetric number [10]. In contrast to node (link) connectivity, these new measures consider both the cost to damage a network and how badly the network is damaged. However, from an algorithmic point of view, it is unfortunate that the problem of calculating these measures for general graphs is NP-complete. This implies that these measures are too costly for use within the confines of complex networks.

Contemporary approaches include, but are not limited to, utilizing link and router information in addition to spectral analysis. Should we consider the former approach, one can immediately calculate the subsequent repercussion: an unfortunate dependence on ISPs for data. ISPs, due to an inherent need to maintain their competitive link, opt out of providing network specific data to researchers. However, assuming ISPs provided all requested data, [8 9] delineate the complexities involved in developing mathematical models. Due to these constraints, there has been a thrust towards utilizing spectral analysis and more specifically, the Laplacian spectrum, as a measure of network robustness. Though notable and innovative in all respects, it will be shown using examples and counterexamples, that these current methods are incapable and insufficient to capture the robustness in complex networks [16, 26].

Strides towards categorizing network topologies resulted in the development of a topological metric system [11, 12, 13, 14, 15, 16, 17]: likelihood [17], expansion [16], resilience [15], performance [17], router utilization [16], assortativity [11], distance [11], spectrum [11], betweenness [11], average node degree, NDD (node degree distribution) and JDD (joint degree distribution). These are the most commonly used metrics. A seasoned veteran in the field of complex networks would concur that one metric is unable to define the structure of a network. However, a collection of "golden" metrics will undoubtedly embody the nature of a network [12]: research is ongoing to obtain these "golden" metrics. Considering all metrics above, this paper seeks to highlight performance and assortativity.

* This paper is best viewed in a color format.

Enlightened by efforts to realize the robustness of Internet models, this study exhaustively targets the heart of network robustness by introducing a necessary metric: elasticity (*E*). Elasticity is defined as the area under the curve of throughput vs. percentage of remaining nodes in a network under attack. From a panoramic perspective, elasticity not only analyzes a topology's adaptability to node and link failures but also captures the overall percentage of flows rerouted under the aforementioned failures. Much research on network robustness starts on a connected network and ends at the point in which this network becomes disconnected [25]. Our study is realistic, proving that performance of disconnected networks cannot be overlooked, and is a significant improvement over some of its predecessors by necessitating solely, the topology of a network in order to determine robustness. Hence, we developed a simple algorithm, which accepts as its input the topology of any network, irrespective of the number of nodes or links, which has the potential to target node removal randomly or selectively in order to determine the scale of operational performance.

We validate the legitimacy of our algorithm by comparing results obtained to that of previously published research. Additionally, we have carried out exhaustive experiments on the most frequently used internet topologies: BGP, traceroutes, Whois, Abilene, as well as computer-generated topologies and will publicize our datasets and algorithm. Finally, we correlate the possible interplay between Assortativity and Elasticity and suggest the possibility of using online social networks to model the Internet.

Our motivation for this research stems from the need to provide network specialists with tools necessary to encapsulate the global performance of any topology, without the overheads in obtaining actual router/link data and specifications. This emulates what was done in [8, 9], without demanding costly computing resources and the use of complex theorems. This not only reduces the time necessary to make an informed decision about an impending network project execution, but also reduces the bureaucratic costs involved with obtaining ISPs' topology data. Thus, ISPs can rest assured that their competitive link is not compromised as their network performance is optimized. Our proposed contribution engenders the following:

1. Defines a metric, Elasticity $0 \leq E \leq 1$, inspired by the Performance metric [8, 9] and develops an algorithm, which for a given network topology, details its adaptability to node and link failures.
2. Validates the legitimacy of our model through comparative analysis
3. Extends the removal of nodes and links to 80% of the total in contrast to 20% in previous work [9]. This extension is necessary to obtain an overall worst case scenario perspective for a topology in question, even in the event of a disconnected network
4. Establishes a possible link between the Assortativity metric [29] and Network Elasticity
5. Explores Internet robustness under random and targeted attacks
6. Proposes the exploration of online social networks to model the Internet

## 2. SPECTRAL ANALYSIS FOR ROBUSTNESS

Research on the performance metric has deviated from the course of utilizing actual router and link specific models [8, 9] to spectral models. The former models have specifically investigated the performance of networks under node removal by comparing the performance of two controversial models of the internet: Scale-free and HOT (Heuristically Optimal Topology). Though accurate, the innate complexity of these models has been a vehicle towards utilizing spectral models to define network robustness. As shown below, spectral analysis though simple, lacks applicability to our definition of robustness. For example, the second smallest Laplacian eigenvalue ($\lambda_2$) has been given significant attention and is generally considered an adequate measure of robustness [15].

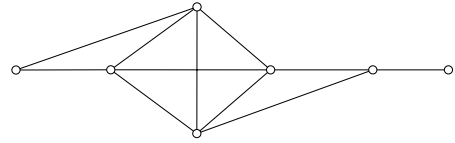

**Figure 1.** Network 1 with a diameter of 4 and $\lambda_2$ = .6766.

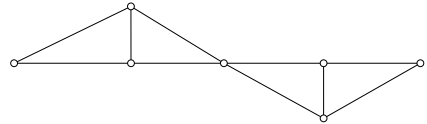

**Figure 2.** Network 2 with a diameter of 4 and $\lambda_2$ = .58579.

Based on the robustness definition presented in the Introduction, network 1 is more robust than network 2. Initially, 42 origin-destination (O-D) flows exist. In network 1, if one node is attacked in such a way that the network becomes disconnected, 20 flows will be delivered, which corresponds to 48% of the initial throughput. In network 2, if the central node is targeted, 12 flows will be delivered, which corresponds to 30% of the initial throughput. Thus, the throughput and consequently the elasticity of network 1 is greater than that of network 2. Likewise, $\lambda_2$ for network 1 is greater than that of network 2. In this case, the second smallest eigenvalue captures our definition of robustness in that the larger the $\lambda_2$, the greater the robustness of a topology. Networks 3 and 4 serve as counterexamples. In network 3, there exist 30 initial O-D flows. If the central node is removed, you can still route 20 flows: 67% of the initial throughput. In network 4, however, there is no central node and in the worse case scenario of a single node removal, out of 306 initial flows 240 will be delivered, which corresponds to 78% of the initial throughput. Thus, network 4 is more robust than the one in network 3. However $\lambda_2$ for network 4 is considerably lower than that of network 3. On this premise, $\lambda_2$ does not consistently capture our definition of robustness.

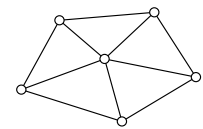

**Figure 3.** Network 3 with a diameter of 2 and $\lambda_2$ = 2.38197.

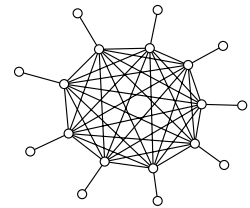

**Figure 4.** Network 4 with a diameter of 3 and $\lambda_2$ = .89023.

From a different approach, researchers in [16] endeavor to use an average of the spectrum of eigenvalues as a more accurate measure of robustness. This works exceptionally but only for small-scale topologies. When dealing with complex networks that span thousands and millions of nodes, finding eigenvalues for all nodes is directly proportional to the resources available in terms of software, processing power and RAM.

## 3. DATA SOURCES

Whois, BGP, Skitter data are frequently utilized Internet topologies obtained from [11]. Inet topology was also obtained from a highly recognized computer topology generator [21]. The Abilene topology, another internet backbone established in 11 US cities, was obtained from [22]. Other computer-generated topologies were obtained from Pajek [23], which is used to analyze large networks, and WHEAT, a C program, which generates lattice topologies. Due to the computational resource requirements of this algorithm, all networks were rescaled using [21] to approximately 1000 nodes.

Rescaling involves characteristics inherent in the d$K$ series where $d$, for the purposes of this study, is a positive integer from 0 to 3 [12]. The computational complexity in rescaling is directly correlated to an increase in $d$. The objective seeks to rescale any given topology to the required number of nodes while preserving topological characteristics. A panoramic overview of this approach details the following:

i. 0$K$: These topologies are produced from average node degree [12, 21]
   - This stochastic technique originated from research done by Erdös and Renyí. However, due to high statistical variance, when applied to higher $d$ values, this approach fails.
ii. 1$K$: Produced from node degree distribution P(k)
iii. 2$K$: Produced from joint degree distribution (JDD)
iv. 3$K$: Produced from links and triangles
   - Utilizes the rewiring technique and can be applied to any of the d$K$ topologies

Researchers from [12] conclude that 2$K$ is sufficient to capture most properties of a given rescaled graph. However, 3$K$ captures all topological characteristics to date. The specific details of the approaches mentioned above are beyond the scope of this paper. However, for the purposes of this research, 2$K$ rescaling was utilized. Further insight can be realized from [12].

## 4. MODEL AND SIMULATION

### 4.1 MODEL

For the subsequent sections of this paper, the following definitions hold true:

- Nodes ($n$): Any interconnecting component on a network: computers, routers, people for social networks
- Links ($m$): A link between nodes
- Graph ($g_i$): The initial graph with $n$ nodes and $m$ links
- Traffic matrix ($T(g_i)$): The initial $n$x$n$ matrix containing each flow (If $j = k$, $T_{jk} = 0$. Otherwise, $T_{jk} = 1$).
- Normalized remaining throughput ($Tp(g)$)
- Bandwidth $B_m = 1$
- ($R(g_i)$): The routing matrix using standard shortest path routing and can be 1 or 0 if a flow traverses an O-D pair
- $X(g_i)$ : A vector formed by stacking all $T(g_i)$ flows
- $f_{max}$: Maximum number of flows through bottleneck link
- $\delta$: Maximum flow for each O-D pair ($\delta = 1/f_{max}$)

Based on our definition of robustness, we introduce a new metric Elasticity ($E$), as the area under the curve of $Tp(g)$ vs. % nodes still functioning in the network. For each network where nodes and links are removed, $Tp(g)$ gives the maximum throughput obtained. Thus:

$$Tp(g) = \max \ \delta T(g) \qquad s.t. \ R(g) X(g) \leq B(g) \qquad (1)$$

Our approach realizes a simplification of this formula where:

$$Tp(g) = \frac{1}{f_{max}} \sum_{jk} T_{jk} \qquad (2)$$

**Algorithm to find Elasticity**

i. Input $g$
ii. Find shortest path and populate $R(g)$
iii. Calculate $f_{max}$
iv. Calculate $Tp(g)$
v. Calculate E, which is the area under the curve of $Tp(g)$ vs. % nodes

### 4.2 RESULTS AND ANALYSIS

Up to this point, discussion about the metric assortativity coefficient ($r$), has been circulated. Assortativity ranges from -1 to 1. In disassortative networks, $r < 0$, the majority of radial links connects nodes of different degrees. The repercussion of this fact indicates that such networks are vulnerable to random failures, targeted attacks, and faster virus dissemination. Conversely, in assortative networks where $r > 0$, the opposite holds true [11]. For all topologies, Table 5, in the appendix, highlights the following key characteristics: assortativity coefficient, number of nodes and links, maximum node degree and average node degree. ORBIS, a rescaling topological tool from [21], provided the capability to rescale networks.

In order to verify ORBIS' rescaling capability, we calculated the metrics mentioned above for the original HOT topology with 939 nodes, a 500, 2000 and finally 5000 node rescaled version. The results are shown in Table 5, in the appendix. Additionally, figure 5 plots Elasticity for the HOT original and rescaled topologies. For all Elasticity graphs, $Tp(g)$ is plotted against the percentage of nodes left in the network. The x-axis has been reversed to portray the initial scenario when all nodes are in the network (100 %) and thus $Tp(g)$ would be 1 (No nodes have been removed). The Legend format for all graphs from figure 9 onwards reads, "Graph name_Assortativity value."

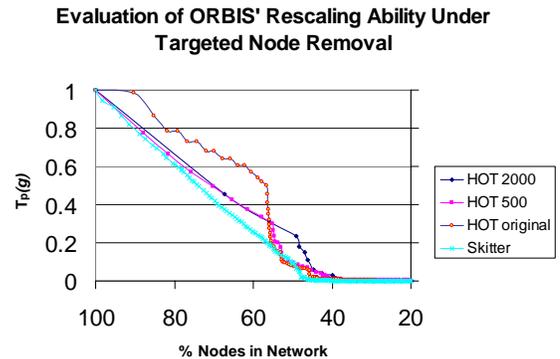

**Figure 5.**

From Table 5, in the appendix, though the number of nodes and links increase, $r$ and the average node degree remain relatively constant. In Figure 5, though the original HOT topology has better performance than the rescaled versions, it can be seen that the rescaled versions still maintain high Elasticity when compared to the Skitter topology, which has one of the highest Elasticity values.

To validate elasticity, we compared our output results to that obtained for the HOT and Scale-free topologies from [20].

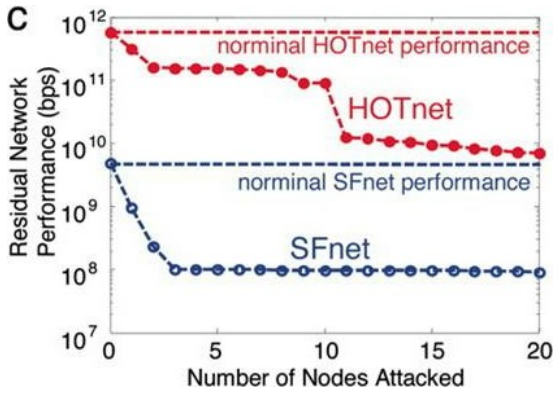

**Figure 6.** [20]

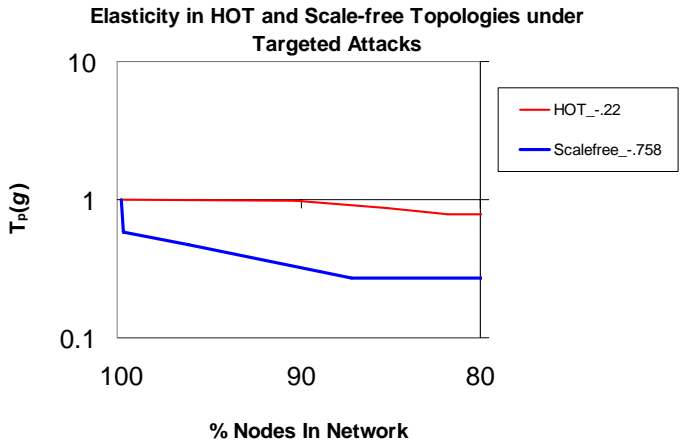

**Figure 7.**

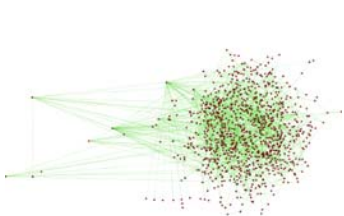 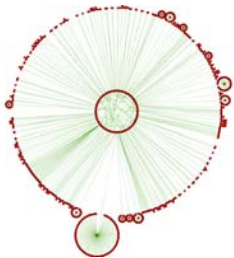

**Figure 8.** Scale-free      **Figure 9.** HOT

Figure 6 shows the performance of HOT (prominent mesh core) compared to Scale-free (high degree hubs) when up to 20% of the high degree nodes were removed. Figure 7 shows our version, relatively comparable to Figure 6. However, our model is based solely on the link-list (Topological structure) of the network and does not include any link or router specific information, as does Figure 6. Yet, our results are highly similar. This result validates the legitimacy of our model.

On this conclusion, we proceed to the subsequent sections of our analysis. For all values of elasticity under random removal, we executed each experiment 20 times and found the average.

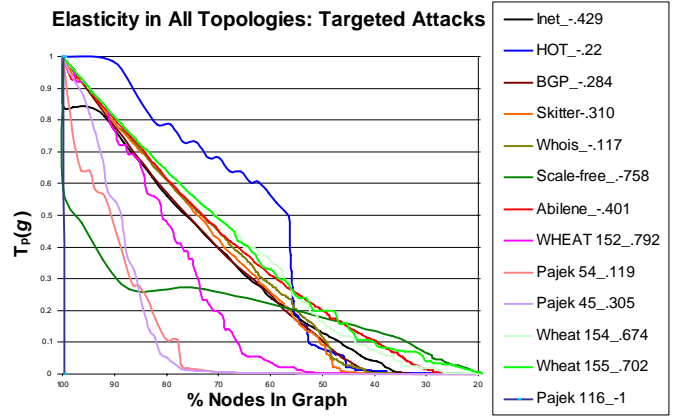

**Figure 10.**

**Table 1.** Elasticity under highest node degree failure

| Graph | Area | Elasticity |
|---|---|---|
| HOT | 35.36124 | 0.4420155 |
| 155 | 31.01468 | 0.3876835 |
| 154 | 30.25828 | 0.3782285 |
| Abilene | 29.87953 | 0.37349413 |
| Whois | 27.3725 | 0.34215625 |
| MySpace | 27.07741 | 0.33846763 |
| Skitter | 26.4573 | 0.33071625 |
| BGP | 25.90381 | 0.32379763 |
| Inet | 25.63658 | 0.32045725 |
| 153 | 22.92724 | 0.2865905 |
| 152 | 19.61782 | 0.24522275 |
| Scale-free | 17.21701 | 0.21521263 |
| 45 | 10.74491 | 0.13431138 |
| 54 | 9.636033 | 0.12045041 |
| 116 | 0.2495 | 0.00311875 |

Table 1 gives the values for elasticity when nodes with the highest node degrees are removed from all topologies in Figure 10. As expected, the HOT topology (Blue trace) which, designed under realistic technological and economic constraints, has the highest elasticity and the Scale-free topology (Green trace) shows poor elasticity. This result is expected due to the fact that the Scale-free topology exhibits few nodes with high degrees. Graph 116, as shown in figure 13, consists of three star-like nodes that interconnect all other nodes. Thus, when these nodes fail, one would expect the entire topology to fail. Topologies 155 and 154, as shown in figures 11 and 12 respectively, are mesh-like topologies. Therefore, the removal of nodes has minimal effect on their elasticity. The next tier of high performing topologies are all Internet topologies, including skitter, Abilene, BGP, Whois. Figures 14, 15, 16 and 17 provide a visual on these topologies. They all have a prominent mesh-like core (comparable to the HOT topology) that provides for high *E*. At this juncture, one should notice the relationship between *r* and *E*. From figure 10, the conclusion can undoubtedly be made that a high percentage of topologies with a negative *r* values, are internet topologies and have the highest *E* values. Conversely, low performing topologies have positive *r* values. This phenomenon will be investigated in later sections.

Previous work compared the removal of 20% of the highest degree (worse case) nodes and as shown in figure 7, the HOT topology had the overall best performance. However, figure 10, taking into consideration the entire network, clearly gives a counter argument to the claim mentioned above. HOT demonstrated excellent Elasticity when 50% of the high degree nodes were removed. However, past this threshold, if more nodes are removed, this topology deteriorates to the worse Internet topological structure. However, Abilene has notably consistent qualities. Its high Elasticity is exhibited consistently under targeted attacks even past the 50% threshold experienced by HOT.

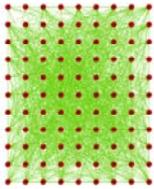
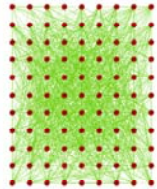

**Figure 11.** Topology 155     **Figure 12.** Topology 154

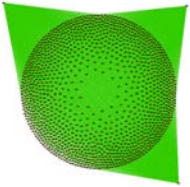
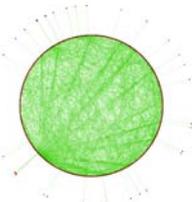

**Figure 13.** Topology 116     **Figure 14.** Whois topology

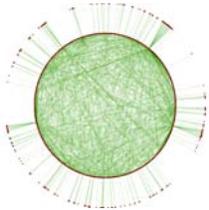
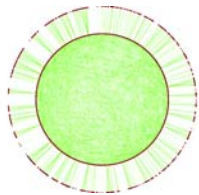

**Figure 15.** BGP     **Figure 16.** Abilene

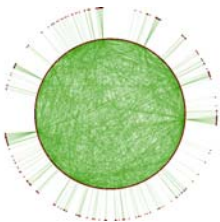

**Figure 17.** Skitter

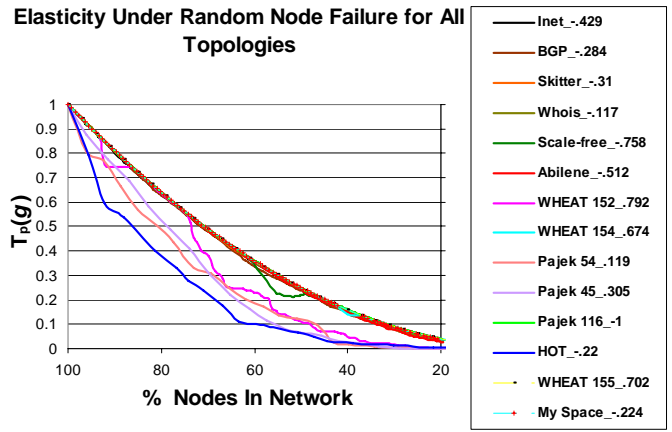

**Figure 18**

**Table 2.** E under random node failure

| Graph | Area | Elasticity |
|---|---|---|
| 116 | 33.4223 | 0.417779 |
| 155 | 33.4058 | 0.417573 |
| 154 | 33.341 | 0.416763 |
| MySpace | 33.3372 | 0.416715 |
| Skitter | 33.1151 | 0.413939 |
| Whois | 33.0163 | 0.412704 |
| Scale-free | 32.8113 | 0.410141 |
| Inet | 32.6983 | 0.408729 |
| BGP | 32.4478 | 0.405598 |
| Abilene | 30.0752 | 0.37594 |
| 152 | 26.6476 | 0.333095 |
| 45 | 23.1055 | 0.288819 |
| 54 | 22.5472 | 0.28184 |
| HOT | 18.3556 | 0.229445 |

Figure 18 shows the response for all topologies under random node attacks. In line with the results in [20], HOT should have been among the best performing network under random node attack. However, as shown above, this claim holds true for the first 5% of nodes removed. As the number of random nodes removed increases, the HOT topology plummets drastically and falls within the sector of graphs with positive $r$ values, the worse performing topologies. Surprisingly, Abilene maintains the highest average Elasticity over 80% for both random and targeted node removals. This rather strange behavior of the HOT topology demonstrates the difficulties involved in computer-generated topologies. One interesting observation is that the Scale-free topology in addition to Graph 116, under random attacks, exhibits elasticity comparable to the other high performing topologies: including Abilene and skitter. One reason is alluded to by the fact that, under random attacks, the probability of removing a highly connected hub is low. Intuitively, one would expect the elasticity of a network under random attack to be higher than that of a network under targeted attack. We observed that all topologies, except HOT adhered to this generality. We surmised that perhaps the structure of the HOT topology with high node degree hubs on the periphery, may have contributed to this behavior. However, we intend to comprehend this topology as part of our future work.

At this point, we attempt to delve deeper into the relationship between *r* and *E* and perhaps formulate some intuitively simple heuristic. The results in figure 19 below make this task impossible based on the limited number of topologies evaluated so far. Consider the example of HOT in figure 19. It has higher *E* than Inet, yet a lower *r* value. Should one attempt to formulate a rule, such that *E* is inversely proportional to *r*, the following counter example would make this heuristic void. Whois has a higher *r* value than HOT. However, its *E* is lower than HOT's *E*. This proves that the relationship is not as clear-cut as one would expect and extensive analysis on numerous topologies must be completed prior to formulating rules.

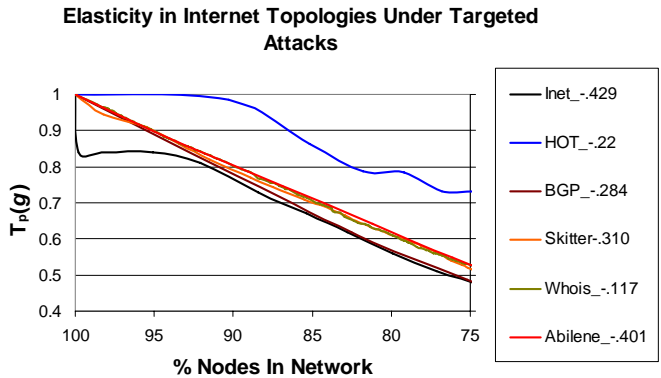

**Figure 19.**

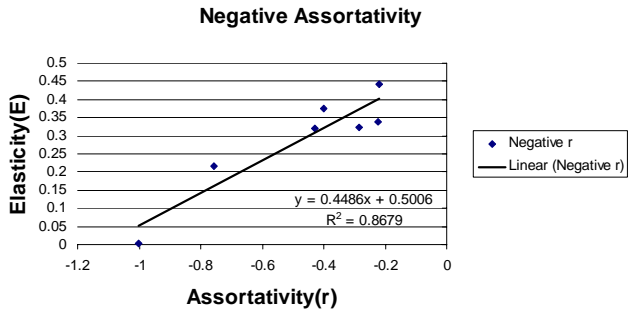

**Figure 20**. E vs r for graphs with negative assortativity, under highest node degree removal

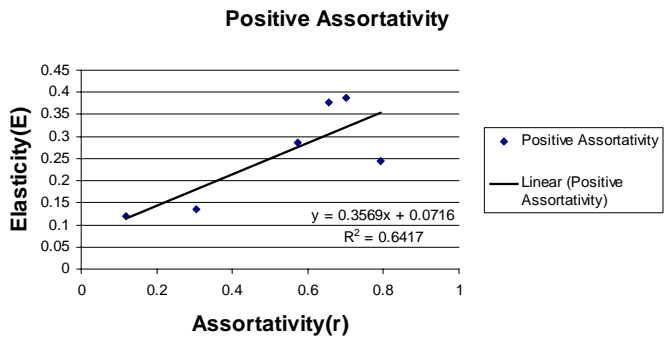

**Figure 21.** E vs r for graphs with positive assortativity, under highest node degree removal

Figure 20 and 21 show a possible correlation between elasticity and assortativity. However, additional graphs must be analyzed to strengthen this correlation. Figure 22 shows the elasticity for all networks under random link removal. Similar conclusions can be drawn for nodes with comparable links: all topologies with a negative *r* value have an exceedingly higher *E* value than those with a positive *r* value. The Whois seemingly has the highest Elasticity value compared to the HOT topology with the lowest. One explanation for this stems from the fact that the Whois topology has the most links, of all topologies evaluated, compared to HOT with the least number of links. Thus, the number of links in a topology will directly affect *E* when links are removed.

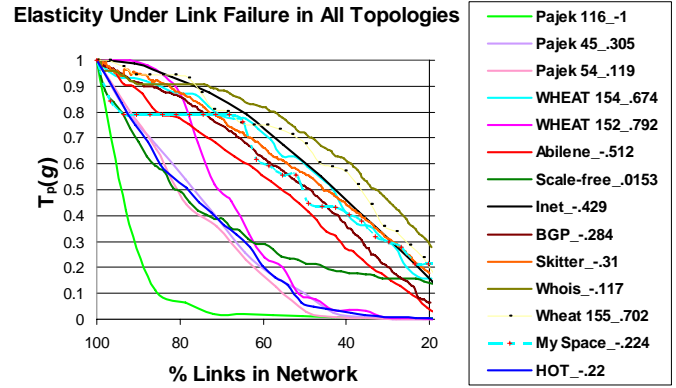

**Figure 22.**

## 5. ELASTICITY IN SOCIAL NETWORKS

MySpace, facebook, orkut, YouTube and Flickr are online social networks. We obtained the MySpace dataset from [14], which included 100,000 nodes and over 6 million links. YouTube and Flicker were obtained from [28]. Once again, due to the extensive computational resources required for such analysis, our team resorted to ORBIS and rescaled these topologies to approximately 1000 nodes. Our objective to establish correlations between social networks and computer networks produces intricate results. Theoretically, both networks are similar and one can hence make the following parallels:

**Analysis of Social Networks (SN) and Internet (IN) Topologies**

**1. Objective of topology analysis:**

   i.  " Understanding structure and mechanism of society"[29](SN)
   ii. Understand structure and mechanism of the internet (IN)
   iii. "Improving spreading of news and opinions"[29](SN)
   iv. Improving spreading of information (data, protocols)(IN)

**2. Tools used in analysis:**

   i.  Weighted network models(SN & IN)
   ii. Undirected or directed links(SN & IN)

**3. Concepts used in analysis:**

   i.  -"Best results are obtained by searching out of your community"[29](SN)
   ii. –Use of proxy server to obtain common information but venture beyond the proxy to obtain new information(IN)
   iii. -Granovetter's Weak Ties Hypothesis ("The relative overlap of two individual's friendship networks varies directly with the strength of their tie to one another"[29](SN)
   iv. -Theory of preferential attachment (Attaching new nodes to existing nodes with high node degree)(IN)

Under both targeted and random node removal in figures 23 and 24 respectively, MySpace exhibited similar *E* as the Abilene network. Should we concur with [20] that the internet remains robust in the face of both random and targeted node failure, we must also concur that

MySpace can be a candidate to represent Internet-like topologies. Thus, there exists the probability that online social networks do exhibit similar characteristics, both theoretically and structurally, to Internet topologies and thus, can perhaps provide valid datasets to model the Internet.

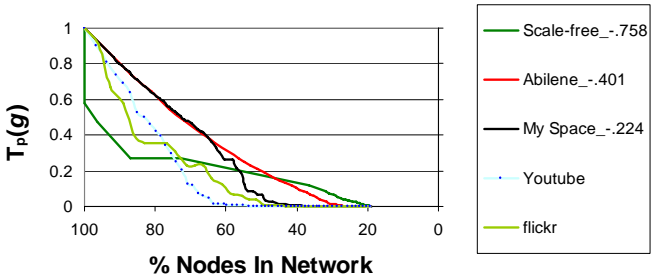

**Figure 23.**

**Table 3.** Elasticity of Social Networks under targeted attacks

| Graph | Area | Elasticity |
|---|---|---|
| Abilene | 29.87953 | 0.373494 |
| MySpace | 27.07741 | 0.338468 |
| Flickr | 17.92365 | 0.224046 |
| YouTube | 17.42864 | 0.217858 |
| Scale-free | 17.21701 | 0.215213 |

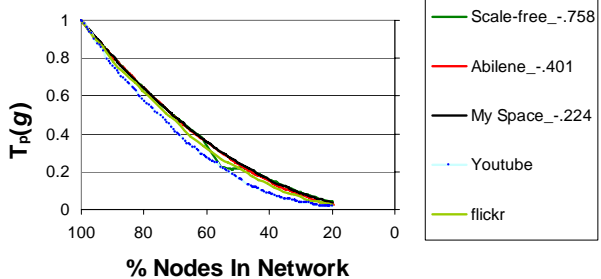

**Figure 24.**

**Table 4.** Elasticity of Social Networks under random attacks

| Graph | Area | Elasticity |
|---|---|---|
| MySpace | 33.3372 | 0.416715 |
| Scale-free | 32.8113 | 0.410141 |
| Flickr | 31.2968 | 0.39121 |
| Abilene | 30.0752 | 0.37594 |
| YouTube | 28.2176 | 0.35272 |

Figures 25 and 26 show the node degree distributions (NDD) for MySpace and Abilene respectively. Both topologies have similar node degree distributions which are comparable to the NDD for general social networks [29]. Additionally, the NDD were plotted for all other Internet graphs, and the graphs exhibited similar NDD. Though it has been proven that different topologies can have similar node degree distributions [19], one must seek to examine all possible topological characteristics to classify topologies.

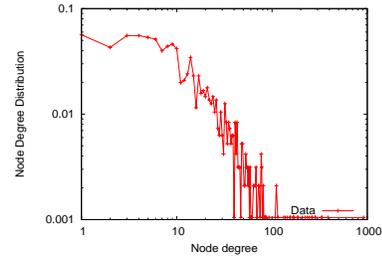

**Figure 25.** NDD for MySpace

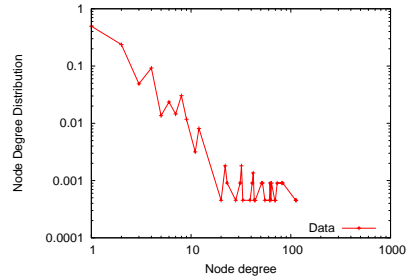

**Figure 26.** NDD for Abilene

From figure 19, one can conclude that the majority of the Internet topologies adhered to the relationship that the lower the assortativity coefficient, the higher the elasticity. This holds true because the topologies cited are all Internet oriented. Their performance decays with the same gradient. This conclusion cannot be made for any of the computer generated topologies as yet. Comparisons with such topologies yield controversial and inconclusive results. To date, a legitimate topology generator, which embodies the characteristics of any particular topology, "remains in the pipeline." However, Abilene performs well under elasticity.

## 6. CONCLUSIONS AND FUTURE WORK

Characterization of network topologies, though recent to the research community, epitomizes the essential concepts inherent in modeling complex systems. In this work, we propose a new metric for robustness-elasticity, which is an approximate predictor of network performance under node and link failures for all topologies. Through comparative analysis our metric has been validated. Unlike previous research [20], which defined performance with reference to actual router and link information, our approach is strictly based on saturation of a homogeneous capacity network. Though strategically novel within the confines of complex networks analysis, our results are relatively accurate and simple to obtain, considering the global topology even in the event of multiple disconnected components. In addition, our work has generated a possible relationship between elasticity and assortativity. Our model explored and concurred with the fact that Internet topologies are less affected by both random and targeted attacks than the Scale-free topologies [20].

Utilizing online social networks to model the Internet seems promising. MySpace demonstrated identical characteristics to the Internet topologies. Therefore, there exists the possibility for MySpace to serve as an Internet model. However, YouTube and Flickr obtained comparable *E* values to Internet topologies under random attacks but portrayed a Scale-free type *E*, under targeted attacks. One reason originates from the class structure of online networks. YouTube and Flickr define a class of online networks where users share photos and videos in contrast to MySpace, where users form purely online social networks.

Elasticity provides several future research avenues. First, algebraic connectivity, though unable to consistently capture the robustness of networks, does provide a novel avenue to capture the robustness of networks utilizing key components of the laplacian spectrum. Secondly, there may exist a relationship between elasticity and assortativity. However, a quantitative approach that necessitates mathematical correlations between the two metrics must be investigated. Finally, in an effort to corroborate the general claim that online social networks can be used to model the Internet, extensive tests encompassing online social networks such as orkut and Facebook are necessary.

## 7. ACKNOWLEDGEMENTS

Special thanks to Alan Mislove and Dr Sue Moon who provided social network databases. This research was partially supported by the Netherlands Organization for Scientific Research (NWO) under project number 643.000.503.

## 8. REFERENCES

[1] H. Frank and I. T. Frisch, "Analysis and design of survivable networks," IEEE Transactions on Communications Technology COM-18, 567, 1970.
[2] D. Bauer, F. Boesch, C. Suffel, and R. Tindell, "The Theory and Application of Graphs,"Wiley, New York, pp. 89–98, 1981.
[3] F. Harary, "On conditional edge-conectivity of graphs," Networks Volume 13, 346, 1983.
[4] A. H. Esfahanian and S. L. Hakimi, "On Computing a Conditional Edge-connectivity of a Graph," Journal of Information processing Letters Volume 27, 195, 1988.
[5] M. S. Krishnamoorthy and B. Krishnamirthy, "Fault diameter of interconnection networks," Computers and Mathematics with Applications 13, 577, 1987.
[6] V. Chv´atal, "Tough graphs and Hamiltonian circuits," Discrete Math Volume 5, 215, 1973.
[7] H. A. Jung, "On a class of posets and the corresponding comparability graphs," Journal of Combinatorial Theory B 24, 125, 1978.
[8] M. Cozzen, D. Moazzami, and S. Stueckle, "Seventh International Conference on the Theory and Applications of Graphs" Wiley, New York, pp. 1111–1122, 1995.
[9] N. Alon, "Eigenvalues and expanders," Combinatorica Volume 6, 83 1986.
[10] B. Mohar, "Isoperimetric numbers of graphs," Journal of Combinatorial Theory Series B 47, 274, 1989.
[11] P. Mahadevan , D. Krioukov, M. Fomenkov, B. Huffaker, X. Dimitripoulos, k. Claffy, A. Vahdat,"Lessons from Three Views of Internet AS-Level Topology," CAIDA Technical Report TR-2005-02.
[12] P. Mahadevan, D. Krioukov, K. Fall, A. Vahdat, "Systematic topology analysis and generation using degree correlations," SIGCOMM, 2006.
[13] P. Mahadevan, A. Vahdat, D. Krioukov, B. Huffaker, K. Claffy,,"Impact of Degree Correlations on Topology Generators," SIGCOMM, 2005.
[14] C. Gkantsidis, M. Mihail, E. Zegura, "Spectral analysis of Internet topologies," IEEE INFOCOM, 2003.
[15] P. Mahadevan, D. Krioukov, M. Fomenkov, B. Huffaker, X. Dimitropoulos, kc claffy, A. Vahdat, "The Internet AS-level topology: Three data sources and one definitive metric," Computer Communication Review, vol. 36, no. 1, 2006.
[16] ] J Wu, Y Tan, H Deng, Y Li, B Liu and X Lv, "Spectral Measure of Robustness in Complex Networks," arXiv:0802.2564, 2008.
[17] H. Tangmunarunkit, R. Govindan, S. Jamin, S. Shenker, and W.Willinger, "Network topology generators: Degree-based vs. structural," ACM SIGCOMM, 2002.
[18] L. Li, D. Alderson, W. Willinger, and J. Doyle, "A first Principles Approach to understanding the Internets router-level topology." ACM SIGCOMM, 2004.
[19] D. Alderson, L. Li, W. Willinger, C. Doyle, "Understanding Internet Topology: Principles, Models, and Validation," IEEE/ACM TRANSACTIONS ON NETWORKING, AFOSR, SIGCOMM, 2005.
[20] J. C. Doyle, D. L. Alderson, L. Li, S. Low, M. Roughan, S. Shalunov, R, Tanaka and W. Willinger, 'The "robust yet fragile" nature of the Internet," PNAS 2005.
[21] P. Mahadevan, C. Hubble, D. Krioukov, B. Huffaker and A Vahdat, "Orbis: Rescaling Degree Correlations to Generate Annotated Internet Topologies," Proceedings of the ACM SIGCOMM Conference, 2007.
[22]Abilene core topology, http://www/stanford.edu /services/internet2/ Abilene.html.
[24] J. F. Kurose, K. W. Ross. "Computer Networking: A Top-Down Approach." Third edition 2005.
[25] Analysis of Topological Characteristics of Huge Online Social Networking Services http://an.kaist.ac.kr/traces/WWW2007.html.
[26] A. Jamakovic and S Uhlig, "Influence of the network structure on robustness," IEEE 2007.
[27] Online Social Networks, www.mpi-sws.mpg.de/~amislove/
[28] J. Kertesz, J Onnela, J Saramaki, J Hyvonen, K Kaski, J Kumpula, D Lazer, G Szabo, A Barabasi, "Social networks from the perspective of Physics," EECS, 2007.
[29] M. E. J. Newman, "Mixing patterns in networks," Physical Review E 67, 026126, 2003.

## 9. APPENDIX

**Table 5.** Common metrics for cited topologies

| Topology | r | n | m | Max Node Degree | Average Node degree |
|---|---|---|---|---|---|
| **HOT_initial** | -.220 | 939 | 988 | 91 | 2.10 |
| **HOT_500** | -.341 | 408 | 444 | 49 | 2.17 |
| **HOT_2000** | -.163 | 1773 | 1936 | 194 | 2.18 |
| **HOT_5000** | -.154 | 4774 | 5229 | 485 | 2.19 |
| **Abilene** | -.401 | 2209 | 3624 | 113 | 3.28 |
| **BGP** | -.284 | 1074 | 2046 | 220 | 3.81 |
| **Skitter** | -.310 | 1158 | 3088 | 303 | 5.33 |
| **Whois** | -.117 | 1169 | 6389 | 260 | 10.93 |
| **My Space** | -.224 | 955 | 10976 | 914 | 22.98 |
| **Inet** | -.429 | 934 | 2222 | 222 | 4.75 |
| **Scalefree** | -.758 | 1000 | 4022 | 743 | 8.044 |
| **Pajek 54** | .119 | 1059 | 1253 | 12 | 2.36 |
| **Wheat 152** | .792 | 1000 | 1975 | 4 | 3.95 |
| **Wheat 153** | .574 | 1000 | 3926 | 8 | 7.852 |
| **Wheat 154** | .674 | 1000 | 9726 | 20 | 19.452 |
| **Wheat 155** | .702 | 1000 | 13558 | 28 | 27.11 |
| **Pajek 45** | .305 | 909 | 1124 | 12 | 2.47 |
| **Pajek 116** | -1 | 1303 | 3900 | 1300 | 5.98 |